\documentclass[aps,pra,showpacs,twocolumn,groupedaddress,nofootinbib]{revtex4}

\newcommand \beq{\begin{eqnarray}}
\newcommand \eeq{\end{eqnarray}}

\newcommand{\feynslash}[1]{{#1\kern-.5em /}}
\def\simge{\mathrel{%
         \rlap{\raise 0.511ex \hbox{$>$}}{\lower 0.511ex \hbox{$\sim$}}}}
\def\simle{\mathrel{
         \rlap{\raise 0.511ex \hbox{$<$}}{\lower 0.511ex \hbox{$\sim$}}}}

\begin{document}

\title{Elementary quantum mechanics of the neutron with an electric dipole moment}

\author{Gordon Baym$^a$ and D.\ H.\ Beck,$^a$}
\affiliation{\mbox{$^a$Department of Physics, University of Illinois, 1110
  W. Green Street, Urbana, IL 61801} 
}
\date{\today}

\begin{abstract}
  The neutron, in addition to possibly having a permanent electric dipole moment as a consequence of violation of time-reversal invariance, develops an induced electric dipole moment in the presence of an external electric field.  We present here a unified non-relativistic description of these two phenomena, 
in which the dipole moment operator, $\vec{\cal D}$, is not constrained to lie along the spin operator.  Although the expectation value of $\vec{\cal D}$ in the neutron is less than $10^{-13}$ of the neutron radius, $r_n$, the expectation value of $\vec {\cal D}\,^2$ is of order $r_n^2$.   We determine the spin motion in external electric and magnetic fields, as employed in past and future searches for a permanent dipole moment, and show that the neutron electric polarizability, although entering the neutron energy in an external electric field, does not affect the spin motion.  In a simple non-relativistic model we show that the expectation value of the permanent dipole is, to lowest order, proportional to the product of the time reversal-violating coupling strength and the electric polarizability of the neutron. 

\end{abstract}




\maketitle


\section*{Introduction}

The neutron is expected to have a permanent electric
dipole moment, as a consequence of the violation of charge conjugation-parity (CP) or time-reversal invariance in nature \cite{theory}.  Searches for this effect have been carried out even before it was realized that CP is violated \cite{jimsmith}, and future experiments are under construction \cite{snsExpt1}.   The current experimental upper limit on the neutron electric dipole moment is $3\times 10^{-26}$ e-cm \cite{ILL}.  The common view of such an electric dipole moment is that it must be permanently aligned along the neutron spin (somehow attached with duct tape), since the spin defines the only axis in the neutron.\footnote{As Salpeter wrote in the context of the electron  \cite{salpeter},  ``A  parity-violating perturbation, corresponding physically to a permanent electric dipole moment of an electron parallel to its spin, is introduced into the Dirac equation for an electron.''} The formal justification is based on Lorentz invariance  -- that the neutron belongs to a spin-1/2 representation of the Lorentz group -- and the Wigner-Eckart theorem.   Within this framework, the microscopic interaction of the neutron with an external electric field is $ -2d_n \vec S \cdot\vec E$, plus relativistic corrections, where $d_n$ is the magnitude of the electric dipole moment, $\vec E$ is the electric field, and $\vec S$ is the spin of the neutron \cite{salpeter,hh,dgh}.  
   
    However, a simple thought experiment reveals that the electric dipole moment of the neutron cannot be irrevocably attached to the spin. Imagine a neutron in a strong magnetic field that effectively locks its spin along the field.  Then apply an electric field to the neutron in an orthogonal direction;   the electric field will induce internal charge separation and thus induce an electric dipole moment parallel to the electric field, $e\langle \vec {\cal D}\rangle = \chi_n \vec E$, where $\chi_n$ is the electric polarizability or susceptibility of the neutron.   Such polarization of the neutron by electric fields is a well-measured effect in Compton scattering of photons on neutrons \cite{schmied,alan,baranov,kossert,drechsel}.   The dipole moment is not generally constrained to lie along the spin.   The total dipole moment of the neutron, the sum of the intrinsic and induced moments, is given by the expectation value of the dipole operator,  $\vec {\cal D}$ -- which  in lowest order is  $\sim\int d^3r \,\vec r\sum_{quarks, i}Q_i\bar q_i(\vec r)\gamma_0 q_i(r)$,
where $Q_i$ is the quark (and antiquark) charge in units of the charge $e$ of the proton, and the $q_i$ are the quark (and antiquark) field operators.  The dipole operator does not distinguish a permanent from an induced dipole moment; rather the existence of a permanent dipole moment is a property of the states of the neutron.

We ask here the microscopic question: {\em what is the non-relativistic quantum mechanical structure of the neutron that gives a non-zero expectation value of the dipole moment in the absence of external fields, and at the same time allows it, when in an external electric field, to have an electric dipole moment not along the spin?}   This question complements the principal focus of previous work on the neutron electric dipole moment, to understand the response of the neutron to externally applied electric and magnetic fields \cite{salpeter}.

    Quite generally,  the electric dipole moment operator  is independent of the nucleon spin, or total angular momentum operator  $\vec  S$, the sum over quark spins, $\sim\frac12 \sum_{quarks,i}\bar q_i \vec\sigma q_i$, plus orbital and gluonic contributions \cite{nspin}; the $\vec\sigma$ are the Pauli spin matrices.
The two operators  $\vec {\cal D}$ and $\vec  S$ obey independent commutation relations, as well as the 
commutation relation $[{\cal D}_i,  S_j] = \epsilon_{ijk} {\cal D}_k$ of a vector with the angular momentum.
The Wigner-Eckart theorem requires only that the {\em expectation value} of the electric dipole moment must lie along the expectation value of the angular momentum in the {\em absence} of external fields.

 We first write the full neutron ground state in terms of the neutron states in the absence of CP-violation.  We denote the (even parity) ground state of the neutron in the absence of CP-violation with $S_z=1/2$ by $ |\uparrow\rangle$.   In order to produce a non-zero (odd parity) dipole moment, the CP-violating interaction, a pseudoscalar,  must be odd in both P and T, and mix in a small component of a relative odd parity excited neutron state, which we denote by $|(t) \uparrow\rangle $, a linear combination of all odd parity, spin-1/2 excited neutron states with $S_z=1/2$; thus the ground state  of a spin-up neutron has the structure, 
\beq
  |N\uparrow\rangle = |\uparrow\rangle + \eta |(t) \uparrow \rangle,
  \label{eta}
\eeq
where $|\uparrow\rangle$ denotes the spin-up ground state of the neutron in the absence of CP violation,  $(t)$ labels the component mixed in by time-reversal violation, and $\eta$ is a small constant which can be taken to be real and positive by suitable choice of phase of $|(t) \uparrow \rangle$.  (See Eq. (\ref{t}) with (\ref{v}) below for the explicit form of $\eta |(t) \uparrow \rangle$ for a particular CP violating interaction.)
The lowest odd-parity spin-1/2 state of the neutron is the N(1535) [and the lowest odd-parity spin-3/2 state is the N(1520)]. 

Similarly the down spin state of the neutron, which is reached by a rotation of $\pi$ about a transverse axis, is 
\beq
  |N\downarrow\rangle = |\downarrow\rangle + \eta |(t) \downarrow \rangle.
  \label{eta-down}
\eeq
On the other hand, if we write the time-reversal operation for spin-1/2 states as $\psi \to i\sigma_2 \psi^*$ (which for example takes  
$|\uparrow\rangle$ to $-|\downarrow\rangle$ and $|\downarrow\rangle$ to $+|\uparrow\rangle$) the time-reversed state of the spin-up neutron must be 
\beq
 {\cal T} |N\uparrow\rangle =- |\downarrow\rangle + \eta |(t) \downarrow \rangle;
  \label{eta-T}
\eeq
 the second term results from an explicitly time-reversal violating interaction, producing the relative minus sign.  In the absence of external fields, states (\ref{eta-down}) and (\ref{eta}) have the same energy as a consequence of rotational invariance.  On the other hand, the energy of the time-reversed state  (\ref{eta-T}), calculated with the time-reversed Hamiltonian, is equal to that of state  (\ref{eta}) calculated with the original Hamiltonian.

  In the absence of external fields, the electric dipole moment of the neutron in the state $|N\uparrow\rangle$ is given to order $\eta$ by
\beq
    \langle \vec d_n \rangle =  e\langle N\uparrow| \vec {\cal D}\,| N\uparrow\rangle  =   e \eta \langle \uparrow|\vec {\cal D} \,|(t)  \uparrow \rangle + c.c.\,.
    \label{EDM}
\eeq
The neutron has spin-1/2; in the absence of applied fields,
\beq
\langle \vec S\, \rangle = \langle N\uparrow| S_z \,| N\uparrow\rangle  = \frac12 \hat z.
\label{spin}
\eeq
Because the spin and electric dipole operators are both rank one (rotational) tensor operators, the expectation values of their components are related by the Wigner-Eckart theorem with
appropriate Clebsch-Gordan coefficients, and thus,
\beq
    \langle \vec d_n \rangle =  d_n\hat z = e\eta  \langle \uparrow |{\cal D}_z|(t)\uparrow \rangle\hat z + c.c..
    \label{EDM1}
\eeq
Note that the expectation value of the dipole moment in the state (\ref{eta-down}) has the opposite sign of that in the state (\ref{eta}), while in the time-reversed state (\ref{eta-T}) the sign is the same as that in the (\ref{eta}).

   The operator ${\cal D}_z$ acting on $|\uparrow\rangle$ is a linear superposition of excited odd parity neutron states of spin-1/2 and spin-3/2, with $S_z$ =1/2, in the ratio  1:-$\sqrt2$ as given by the Clebsch-Gordan coefficients.  These two states, which are not themselves eigenstates of the unperturbed Hamiltonian but rather are superpositions of unperturbed neutron states of all energies, are defined by
\beq
  {\cal D}_z |\uparrow\rangle = \frac13\langle \vec{\cal D}\,^2\rangle^{1/2}\left(\sqrt2\, |(d) 3/2\rangle- |(d)1/2\rangle\right),
  \label{clebschd}
\eeq
where $(d)$ labels the states produced by the dipole moment operator, and $\langle\vec{\cal D}\,^2\rangle$ is the expectation value of the square of the dipole moment in the ground state of the neutron.\footnote{
One can also write the state (\ref{clebschd}) in terms of reduced matrix elements as 
\beq
   {\cal D}_z |\uparrow\rangle = \sqrt\frac23\, \langle (d)3/2|| {\cal D} || 1/2\rangle \,|(d) 3/2\rangle     -\sqrt\frac13  \langle (d)1/2|| {\cal D} || 1/2\rangle  \,|(d)1/2\rangle; \nonumber
\eeq
here the reduced matrix elements obey
$
  \langle (d)3/2|| {\cal D} || 1/2\rangle = \langle (d)1/2|| {\cal D} || 1/2\rangle =  \frac1{\sqrt3}\langle \vec{\cal D}\,^2\rangle^{1/2}. 
$
Note that if one takes only the lowest odd parity neutron states the two reduced matrix elements would not be equal.}
  (For simplicity, we suppress further writing the $S_z$ value in the states, where it is obvious.)  To derive the normalization in Eq.~(\ref{clebschd}), we note that in the ground state, $ \langle {\cal D}_i  {\cal D}_j\rangle  = \delta_{ij}\langle\vec{\cal D}\,^2\rangle /3$, which follows from the fact that the rank two tensor operator ${\cal D}_i {\cal D}_j$ has only angular momentum $J$ = 0 and 2 components, and the expectation value of the $J$ = 2 component between spin--1/2 states must vanish; the $J=0$ component is $\delta_{ij} \vec{\cal D}\,^2/3$.   Thus
\beq
  \langle d_n \rangle =   \frac13 e\eta   \langle (d)1/2\uparrow|(t)\uparrow\rangle   \langle\vec{\cal D}\,^2\rangle^{1/2} + c.c. ;
    \label{EDM2}
\eeq
this equation relates the permanent electric dipole moment to the CP violation parameter $\eta$ and the root mean square dipole moment in the neutron.  As we see below, $\langle\vec{\cal D}\,^2\rangle$ is measured by the electric polarizability of the neutron.

\section*{Response to static external electric fields}

  In the presence of a weak external electric field, $\vec E$, the state of the neutron becomes
\beq
  | (E)N\uparrow  \rangle =    | N \uparrow \rangle  +  e{\cal V} \vec{\cal D}\cdot\vec E\, |N \uparrow \rangle,
\label{ae}
\eeq
where $(E)$ denotes the presence of the electric field, the operator
\beq
  {\cal V} = \sum_{n\ne \uparrow} \frac{|n\rangle \langle n |}{\omega_n}
  \label{v}
\eeq
commutes with the total angular momentum, 
and $\omega_n$ is the energy of the excited neutron state $|n\rangle$ measured with respect to the neutron mass.  Only odd parity states
survive in the sum in Eq.~(\ref{ae}).   In the absence of time-reversal violation, the state (\ref{ae}) reduces to $ |\uparrow \rangle  +  e{\cal V} \vec{\cal D}\cdot\vec E\, | \uparrow \rangle$.  Under time reversal this state goes into $- |\downarrow \rangle  - e{\cal V} \vec{\cal D}\cdot\vec E\, | \downarrow \rangle$; the odd-parity admixture has the opposite behavior under time reversal as that in the state (\ref{eta}).

The total dipole moment of the neutron is then\\ $\langle d_{tot,i}\rangle =  e\langle (E)N\uparrow | {\cal D}_i\, |(E)N\uparrow \rangle =  d_n\hat z_i +\chi_{n,ij} E_j$,
where the polarizability tensor is given by $\chi_{n,ij} = 2e^2  \langle \uparrow| {\cal D}_i {\cal V} {\cal D}_j |\uparrow\rangle$.
As with ${\cal D}_i {\cal D}_j$, 
only the $J=0$ component, $\vec{\cal D}\cdot{\cal V} \vec{\cal D}\delta_{ij}/3$, of the tensor operator ${\cal D}_i {\cal V} {\cal D}_j$,  has a non-zero expectation value in the neutron ground state, so that $\chi_{n,ij} =\delta_{ij} \chi_n$,  and thus
\beq
  \langle \vec d_{tot}\rangle  =  d_n\hat z +\chi_n \vec E.
 \label{vecd2}
\eeq

    Using Eq.~(\ref{clebschd}), we see explicitly that
\beq
   \chi_n =  2e^2  \langle \uparrow| {\cal D}_z{\cal V} {\cal D}_z |\uparrow\rangle = 
  \frac{2e^2}{9}\langle  \vec{\cal D}\,^2\rangle \left(  \frac1{\omega_{1/2}}+ \frac2{\omega_{3/2}} \right). 
 \label{alpha}
\eeq
Here $1/\omega_{1/2} \equiv  \langle (d) 1/2 | {\cal V} | (d) 1/2 \rangle$, and \\ $1/\omega_{3/2} \equiv  \langle (d) 3/2 | {\cal V} | (d) 3/2 \rangle$. 
The odd parity component of the neutron ground state produces corrections of order $\eta$, and can be neglected.

    The reader may worry at this point how the electric field can produce a dipole moment not necessarily parallel to the spin 
(cf.~Eq.~(\ref{vecd2})).  Clearly the admixture of a spin-3/2 component in the state (\ref{clebschd}) permits the neutron to be in a multi-axis configuration.  Imagine, however, that the energy of the spin-3/2 states entering $\cal V$ are infinitely high, so that the state in the presence of $\vec E$ and $\eta$ is pure spin-1/2. Must the induced dipole moment in this situation be parallel to the spin?  The answer is no; because a transverse electric field mixes in $S_z=-1/2$ components of the excited states, the neutron is no longer in an eigenstate of $S_z$, and hence its spin and induced electric dipole moment need not point in the same direction.   More generally, a transverse electric field can rotate the spin, the transverse components begin at order $|\vec E|^2$.
Note that the induced electric dipole moment can exactly cancel the permanent dipole moment for a suitable choice of electric field.

    Experimentally \cite{schmied,alan,baranov,kossert}, $\chi_n = 1.26 \times 10^{-3}$ fm$^3$; equivalently,  $\chi_n E = 0.86 \times 10^{-31}E_4$ e-cm, where $E_4$ is the applied field in units of $10^4$V/cm.    Even with a static laboratory electric field as large as $10^6$ V/cm, the induced dipole moment is some 300 times smaller than the current experimental upper bound on the permanent dipole moment of the neutron \cite{ILL}.
 To estimate $\langle\vec{\cal D}\,^2 \rangle$ in the ground state of the neutron, we consider only the two lowest excited states, N(1520) and N(1535), in Eq.~(\ref{alpha}), for which the excitation energies are nearly equal, and find
\beq
  \langle\vec{\cal D}\,^2 \rangle \simeq 0.77 \,{\rm fm}^2.
\eeq
  Even though the expectation value of $\cal D$ in the neutron, in the absence of an external electric field, is $< 10^{-12}$ fm, the expectation 
value of its square is on the order of the nucleon radius squared.   The permanent dipole moment of the neutron is
the manifestation of the internal dipole having a difference of probability of pointing in the direction of the spin, as compared to the direction opposite, of less than one part in $10^{12}$.

    Were we to regard the internal dynamics of $\vec{\cal D}$ as harmonic with frequency $\omega$ (of order $\omega_{1/2}$) and effective mass $\mu$, then since in the ground state of an oscillator $\langle {\cal D}^2 \rangle =3/2\mu\omega$ we would 
estimate the effective mass associated with the dipole moment to be $\sim$ 130 MeV, approximately the reduced constituent mass of 
a pair of quarks.

\section*{Time-dependent response to external magnetic and electric fields}
  
     Since the spin dynamics form the basis of past and proposed schemes to detect the permanent electric dipole moment of the neutron, we turn next to the equation of motion of the spin, The interaction of a neutron with external magnetic and electric fields, $\vec B$ and $\vec E$, is described by the Hamiltonian
 \beq
   H_{em} =  - \vec\mu_n\cdot \vec B -e\vec{\cal D}\cdot \vec E,
\label{eqn:Hem}
\eeq    
where $\mu_n$ is the magnetic moment of the neutron.
Rotational invariance of the internal Hamiltonian of the neutron implies the equation of motion of the spin in the external
fields,
\beq
\frac{d}{dt}{\vec S} =  -i[\vec S, H_{em}] = 2\mu_N \vec S\, \times \vec B + e \vec{\cal D} \times \vec E,
\label{eq:dsdt}
\eeq
where $\mu_n = |\vec \mu_n|$;
in the final term we use the commutation relation $[S_i, {\cal D}_j] =  i\epsilon_{ijk}{\cal D}_k$, of the angular momentum with a vector.
Taking expectation values of both sides in the neutron ground state, we find
\beq
\frac{d}{dt}{\langle \vec S\rangle} = 2\mu_N \langle \vec S\,\rangle \times \vec B +  \langle \vec d_{tot} \,\rangle \times \vec E,
\label{precess}
\eeq
with $\langle \vec d_{tot} \,\rangle$ given by Eq.~(\ref{vecd2}).   The induced dipole moment vanishes in the cross product,
so that\footnote{\label{F}  Note however that in detection -- via a spin-precession experiment -- of an intrinsic dipole moment of a particle with spin greater than 1/2,  the 
induced dipole moment can also drive the spin; the electric polarizability,  $\chi_{ij}$, need not be isotropic  and thus the
induced moment need not be parallel to the applied electric field (quadratic Stark effect).  The operator $ {\cal D}_i {\cal V} {\cal D}_j$ entering the electric
polarizability has both $J=0$ and $J=2$ tensor components, and in a higher spin state $|J'M'\rangle$, the $J=2$ component gives a non-zero traceless contribution to  $\chi_{ij}$; then rotational invariance about the spin quantization axis, $\hat z$,  implies that the polarization tensor is diagonal with $\chi_{xx}=\chi_{yy} \equiv \chi_\perp \ne \chi_{zz}$.  The induced dipole moment contribution to Eq.~(\ref{eq:dsdt}) is \break
$(\chi_{zz}-\chi_\perp) E_z (\hat z \times \vec E)$.
This driver of spin precession, quadratic in $E$ as expected, vanishes only with perfect alignment of the $\vec E$ and $\vec B$ fields.  Assessing its effects depends on detailed calculation of the asymmetry of $\chi$ and the particular experimental setup.     }
\beq
\frac{d}{dt}{\langle \vec S\,\rangle} = 2\langle \vec S\,\rangle \times (\mu_n  \vec B +  d_n \vec E),
\eeq
yielding harmonic precession of the neutron spin at frequency $\omega_{prec} = 2\left( \mu_n B_z + d_n E_z \right)$.  This dependence of the precession frequency on the electric field is used in searching for the permanent electric dipole moment of the neutron.  Although the precession frequency is independent of the induced moment, the interaction energy, Eq.~(\ref{eqn:Hem}), depends on both the permanent and induced electric dipole moments.  
In a static electric field the energy of a neutron is modified by $\Delta E = -d_n\langle \vec S\,\rangle \cdot \vec E -\frac12 \chi_n E\,^2$; variation of the energy with the field has been used to calculate both permanent dipole moment \cite{dip-lattice} and the polarizability on the lattice \cite{pol-lattice}.

\section*{Non-relativistic model for CP violation}

The underlying physics of the neutron electric dipole moment can be illustrated in a simple, non-relativistic phenomenological interaction in terms of the spin and dipole moment operators:
\beq
  H_{CPV,N} = -a\vec S \cdot \vec{\cal D},
\label{a}  
 \eeq
where $a$, a real constant, has dimensions of force, and is bounded above in magnitude by $a < 10^{-11}$ fm$^{-2}$.   Under time reversal, $\vec S \cdot \vec{\cal D}$ changes sign.  This interaction is analogous to the parity-violating simulated spin-orbit interaction $\sim \vec \Sigma \cdot \vec p$ in cold atom systems, where $\vec \Sigma$ is the atomic hyperspin and $\vec p$ the atomic momentum \cite{spin-orbit}, as well as to the parity-violating nucleon chiral magnetization $\sim \vec S \cdot \vec p$, where $\vec S$ is the nucleon spin and $p$ its momentum, that gives rise to the anapole moment of 
heavy nuclei \cite{bouchiat}.  This phenomenological interaction allows us to give a unified description of both the permanent electric dipole moment and electric polarizability of the neutron.

In the presence of the interaction (\ref{a}), but in the absence of external electric and magnetic fields, the state of a spin-up 
neutron is 
\beq
    |N \uparrow \rangle =   \left(1  + a {\cal V}\vec{\cal D}\cdot \vec S\,\right)|\uparrow\rangle,
 \label{t}   
\eeq
providing an explicit realization of the CP-violating term in the neutron wave function (\ref{eta}),
\beq
     \eta |(t) \uparrow\rangle = a {\cal V}\vec{\cal D}\cdot \vec S\,|\uparrow\rangle.
  \label{etaa}
\eeq
Under time reversal, the state $|N\uparrow\rangle$ goes to  $\left(-1  + a {\cal V}\vec{\cal D}\cdot \vec S\,\right)|\uparrow\rangle$, as in (\ref{eta-T}).

To lowest order in $a$, the permanent dipole moment of the neutron is then
\beq
 \langle \vec d_n\,\rangle &=& a e\langle \uparrow| \vec{\cal D}  {\cal V} \vec{\cal D}\cdot \vec S |\uparrow\rangle\, + c.c.
\eeq
The operator $\vec{\cal D}\cdot \vec S$ acting on the state $ |\uparrow\rangle$ produces a spin-1/2 neutron state of opposite (negative)
parity, which can be written in terms of the states defined in Eq. (\ref{clebschd}) as 
\beq
   \vec{\cal D}\cdot \vec S |\uparrow\rangle  = \frac12 \langle \vec{\cal D}\,^2\rangle^{1/2}\,   |(d)1/2\uparrow\rangle;
 \label{ds}  
\eeq 
then
\beq
 \vec d_n =\frac{a e}{3\omega_{1/2}}\langle \vec{\cal D}\,^2\rangle \hat z.
   \label{dnn}
\eeq

  The permanent electric dipole moment, as this model shows, is closely related to the electric polarizability, $\chi_n$, of the neutron.   Taking the ratio of Eqs.~(\ref{dnn}) and (\ref{alpha}), we find 
\beq
  \frac{d_n}{\chi_n} = \frac{3a}{2e}\frac{1}{1+2\omega_{1/2}/\omega_{3/2}}.
\eeq  
If we again consider only the two lowest excited states, N(1520) and N(1535),  the ratio simplifies to
\beq
  \frac{d_n }{\chi_n}\simeq \frac{a}{2e}.
  \label{ratio}
\eeq

   Although we have not written an explicit coupling of $\vec S$ and $\vec E$ in the presence of a permanent electric dipole of the neutron, as has been used since Salpeter's seminal paper \cite{salpeter},  the spin-electric field coupling arises in the present model as the cross terms in the energy computed to second order in  $-e\vec{\cal D}\cdot \vec E$ and $-a\vec S\cdot\vec{\cal D}$; the effective Hamiltonian for this process is 
  \beq
  \delta H = -ea\vec E\cdot \vec{\cal D}  {\cal V} \vec S \cdot  \vec{\cal D}  +h.c. \to -d_n  \hat S \cdot\vec E,
\eeq
with expectation value in the ground state of the neutron,
\beq
 -ea\vec E\cdot \langle \vec{\cal D}  {\cal V} \vec S \cdot  \vec{\cal D}\rangle  +c.c. = -d_n  \langle \hat S \rangle\cdot\vec E.
\eeq 

  To relate the coupling $a$ in this model to $\eta$, we use Eqs.~(\ref{etaa}) and (\ref{ds}) to write,
  \beq
  \eta = a \langle (t)\uparrow | {\cal V} \vec{\cal D}\cdot \vec S|\uparrow \rangle = \frac{a}{2}\langle {\cal D}^2\rangle^{1/2} \langle (t)  | {\cal V}|(d)1/2\rangle;
\eeq
a first estimate is that  $\langle(t)  | {\cal V}|\uparrow (d)1/2 \rangle \sim 1/\omega_{1/2} \sim (1/3)$ fm, and thus
$a \sim 7 \eta$ fm$^{-2}$.

 \section*{Discussion}
   
  The manifestation of a permanent electric dipole moment in the structure of the neutron will be an increasingly important clue to physics beyond the standard model.   As we stressed here the same dipole operator underlies both the permanent dipole moment and the polarizability.    The short distance physics generating the CP violation is filtered through the dipole operator; thus as one tries to interpret a new round of neutron electric dipole moment measurements, the polarizability and permanent dipole moment must be understood within a common framework.  
  
       An important question, both practical and of principle, is whether and how one can experimentally distinguish a permanent dipole moment from an induced dipole moment.      In an external electric field, the energy of a permanent dipole moment is always linear in the field, while that from an induced moment is quadratic in the field, as discussed above, indicating that one can in principle distinguish the two moments on the basis of their behavior in an applied field.   However, in more complex systems, the distinction between the two moments can be blurred.   While the spin precession of a neutron in an external electric field is independent of the induced dipole moment, the precession of particles with spin greater than 1/2 can depend on the induced moment, as discussed in footnote \ref{F}, and contribute to the spin precession in Eq.~(\ref{precess}).  Although the contribution of the permanent moment to the precession frequency remains linear in the field, while that from the induced moment would be quadratic, whether one can control the electric field adequately in reversing its direction to distinguish the two contributions in practice is not immediately apparent.  
A further complication arises if the particles are not in an angular momentum eigenstate, e.g.,  a complex protein frozen in a parity violating state, and thus can  have a ``permanent'' electric dipole moment which would give rise to a linear dependence of the energy in an external electric field.  
  
   The present examination of the structure of the neutron as a polarizable quantum mechanical system with a permanent electric dipole moment raises a number of issues for further investigation.  
  The first is how to relate the phenomenological description given here to more microscopic physics of CP violation.   As an example, the question remains open of how to relate the dipole moment within  phenomenological quark models of the neutron, e.g., bag models,  to the description of CP violation via the QCD $\theta$ term 
\beq
 L_\theta = \theta \frac{g^2}{32\pi^2}\int d^3x F_a^{\mu\nu}(x) {\tilde F}_{\mu\nu}^a(x);
\eeq  
here $F_a^{\mu\nu}  = \partial^\mu A_a^\nu - \partial^\nu A_a^\mu - gf^{abc} A_b^\mu A_c^\nu$  is the color electromagnetic field tensor of quantum chromodynamics, $A^a_\mu$ is the gluon field with color index $a$, $g$ the QCD coupling constant, and ${\tilde F}^a_{\mu\nu}=\epsilon_{\mu\nu\lambda\sigma}F_a^{\lambda\sigma}$ is the dual field tensor.   One phenomenological approach is to replace  $L_\theta$ by a simple complex quark mass term $me^{2i\theta\gamma_5}$ \cite{baluni,miller}.  This approach however is inadequate, since the complex phase can be eliminated by a chiral rotation, except for contributions involving the chiral anomaly \cite{aoki,bbh}.   The same issue can be present in calculating the dipole moment via lattice gauge theory \cite{sharpe}, but see  \cite{dip-lattice}.

    A second issue is to extend the present phenomenological picture to the proton and to atomic nuclei, for which there are also ongoing searches for a permanent dipole moment \cite{theory}.   An immediate question is to extract the electric polarizability from Compton scattering experiments, including radiative corrections \cite{brown-feynman,LT,kaiser}, which vanish for the neutron.  One can also ask to what extent this picture can be generalized to the electron which, although not composite, has a polarizable screening cloud?

    In summary, we have related the permanent dipole moment of the neutron to an admixture of a component of the neutron wave function with opposite parity and time-reversal symmetry, and shown that the electric polarizability of the neutron involves a similar admixture, but with opposite time-reversal symmetry, in the wave function.  In the absence of external fields, the expectation value of the electric dipole moment lies in the direction of the expectation value of the spin; in an applied electric field the induced moment is parallel to the field.  Although the energy of the neutron in an electric field depends on its polarizability, the precession of the spin in external electric and magnetic fields is independent of the polarizability.  
We have also examined a simple model of the source of the CP-violating component of the particle wave function, showing that there is an approximate cancellation of structure effects in the ratio of the permanent moment and the polarizability.  Such structure effects remain important in interpreting future experiments in terms of fundamental interactions.

\acknowledgments{
 We are grateful to B. Filippone, T. Hatsuda, A. J. Leggett, A. Nathan,  C. J.  Pethick, 
 and J. Shelton for very helpful discussions.  This research was supported in part by NSF Grants PHY-1506416 (to DHB) and PHY-1305891 (to GB), and was partially carried out at the Aspen Center for Physics, supported in part by NSF Grant PHY-1066293.
We thank both referees, W. Haxton and T. Sch\"afer, for their constructive insights.  
}



\end{document}